\documentstyle[prl,aps,multicol,psfig]{revtex}
\begin{document}
\draft
\title{Langevin equation approach to granular flows in narrow pipes}
\author{Tino Riethm\"uller, Lutz Schimansky--Geier, Dirk Rosenkranz, Thorsten P\"oschel%\cite{bylineLSG}
}
\address{Humboldt-Universit\"at zu Berlin, Institut f\"ur
  Physik, \\ Invalidenstra\ss e 110, D-10115 Berlin, Germany}
%\date{\today} 
\date{April 3, 1995}
\maketitle
\begin{abstract}
  The flow of granular material through a rough narrow pipe is
  described by the Langevin equation formalism. The stochastic force
  is caused by irregular interaction between the wall and the granular
  particles. In correspondence with experimental observations we find
  clogging and density waves in the flowing material.
\end{abstract}
% 81.05.Rm Porous materials; granular materials
% 02.50.-r Probability theory, stochastic processes, and statistics 
%          (see also 05 Statistical physics)
% 02.50.Ey Stochastic processes
% 25.75.Ld Collective flow
% 47.55.-t Nonhomogeneous flows
% 83.10.Hh Flow of solids
% 83.50.-v Deformation; material flow
% 83.10.Pp Particle dynamics
\begin{multicols}{2}
When granular material flows through a narrow vertical pipe one
observes recurrent clogging and density
waves~\cite{SchickVerveen,Poeschel,Raafat}. This effect is well known to
physicists and engineers, usually it is undesirable and causes
technological problems, e.g.~in chemical engineering. Density waves
play a major role in the behavior of granular materials and have
been investigated by many authors using various methods: 

Ristow and Herrmann~\cite{RistowHerrmann} reproduced the density
fluctuations in an out-flowing hopper by molecular dynamics that have
been observed experimentally before~(e.g.~\cite{Baxter}) using the
X-ray technique. Baxter and Behringer~\cite{BaxterBehringer} simulated
the flow using an cellular automaton.  Peng and Herrmann~\cite{Peng}
studied a Lattice Gas Automaton~\cite{Frisch} for the flow of granular
material. Using phenomenologically plausible rules for the interaction
of particles and of particles with the wall they could reproduce
density fluctuations which spectrum obeys a power law. Lee and
Leibig~\cite{LeeLeibig} applied the kinetic wave
approach~\cite{Lighthill} to the flow of granular particles through a
pipe. They treated initial random density fluctuations as a set of
distinct homogeneous density regions and considered the motion of the
interfaces between them. Finally they could show, that the evolution
of such a simple model leads to the formation of clusters with a high
density contrast.

The aim of the present paper is to provide a model for the flow of
granular material in a vertical narrow pipe using the
Langevin-equation approach of stochastic forces. Such a approach was
also used by Mehta~et~al.~\cite{MehtaNeedsDattagupta} to describe the
relaxational behavior of a granular pile submitted to vibration and
was proved to be suitable. We will show that our model is able to
reproduce the experimental observations~\cite{SchickVerveen,Poeschel}.
The simulation of the resulting density equation requires much less
computational effort than the direct simulation using molecular
dynamics~\cite{Poeschel,Lee}. We discuss the instability of the
homogeneous flow in the hydrodynamic approximation and provide
critical values for the occurrence of clogging and density waves.

When sand flows through a narrow pipe we assume that there is a
permanent random interaction of the sand particles with the wall of
the pipe. The equations of motion for a single particle which is
subjected to gravity $g$ in positive $x$-direction and which does not
interact with other particles in the low density regime read
\begin{mathletters}
\begin{eqnarray}
  \label{LangevinEq}
    \dot{x}_i &=& v_i\\
    m \dot{v}_i &=& m g - \gamma v_i + \sqrt{2\epsilon\gamma}\,\xi_i(t) \,.
\end{eqnarray}
\end{mathletters}
\noindent The friction $\gamma$ and the Langevin fluctuation term results from
the interaction of the grain with the wall. (Although our model does
not include the interaction of the grains with the air inside the
pipe, in a very simple approximation one can assume that the
fluctuation term accounts for this interaction too.) For the
stochastic force we assume Gaussian white noise ($\langle
\xi_i(t)\xi_j(t+T)\rangle = \delta_{ij} \delta(T)$). The path of a
particle is scattered independently at different places during its
motion downwards which is described by independent impacts in time.
Hence, after the relaxation time $m/\gamma$ the velocities of the
particles obey a Maxwellian distribution with mean value
$v^0=mg/\gamma$.

We apply the collision integral proposed by Prigogine and
Herman~\cite{Prigogine} for the description of particle interactions.
In their investigation it was intended to model vehicular traffic flow
and it has been pointed out by several
authors~(e.g.~\cite{Lee,Poeschel}) that traffic flow on
one-lane highways reveals striking similarities to granular flow in a
pipe.  When a fast moving particle $i$ collides with a slower one $j$
after the collision both grains move with the lower velocity $v_j$.
$$
\bigcirc \stackrel{v_i}{\longrightarrow}\bigcirc\stackrel{v_j}{\rightarrow}
\quad \Rightarrow \quad \bigcirc\bigcirc\stackrel{v_j}{\rightarrow}
$$ 
During the impact the momentum balance of the particles is not
conserved. Due to the strong interaction between the particle and the
wall we suppose that the lost part of the momentum will be taken
over by the wall which is assumed to be coupled with a reservoir.
Introducing the effective cross section $C$ which depends on the
geometry of the pipe and the particles the collision integral reads
\begin{eqnarray}
    \label{Stoss}
    \left(\frac{\partial P}{\partial t}\right)_{S} &\sim&
    \int_{-\infty}^{\infty}  P(x,v,t)~P(x,v^\prime,t)~\left(v^\prime-v\right)~dv^\prime\nonumber\\
    &=& C~P(x,v,t)~n(x,t)\left(u(x,t)-v\right)~,
\end{eqnarray}
where $P(x,v,t)$ is the one-particle probability density in phase
space. The particle density $n(x,t)$ and the mean velocity $u(x,t)$
at position $x$ and time $t$ are given by
\begin{mathletters}
  \begin{eqnarray}
  \label{Stoss2}
  n(x,t)&=&\int_{-\infty}^{\infty} P(x,v,t)~dv\\
  u(x,t) &=& \frac{1}{n(x,t)} \int_{-\infty}^{\infty}v~P(x,v,t)~dv~.
  \end{eqnarray}
\end{mathletters}
\noindent Hence we find the kinetic equation
\begin{eqnarray}
  \label{FPE}
  \frac{\partial P}{\partial t} &+& \frac{\partial}{\partial x}[vP] +
  \frac{\partial}{\partial v}\left[\left(g - \frac{\gamma}{m}v\right) P\right] \nonumber \\
  &=& \frac{\epsilon\gamma}{m^2}\frac{\partial^2 P}{\partial v^2} + CP(x,v,t)n(x,t)(u(x,t) - v)
\end{eqnarray}
and in a homogeneous stationary flow ($n=n^0$, $u=u^0$) 
\begin{equation}
  \label{homogen}
  P^0(v) = \sqrt{\frac{m}{2\pi~k_BT^0}} n^0
  \exp\left(-\frac{m}{2~k_B T} \left(v-u^0\right)^2\right)~.
\end{equation}
\noindent The mean velocity 
\begin{equation}
  \label{uhomogen}
  u^0=\frac{m g}{\gamma} - \frac{C k_B T^0 n^0}{\gamma}
\end{equation}
\noindent depends on the density and the mean square displacement of the velocity:
\begin{equation}
  \label{Tallg}
  T(x,t) = \frac{m}{k_Bn(x,t)}
    \int_{-\infty}^{\infty}\left(v-u(x,t)\right)^2 P(x,v,t)~dv \,.
\end{equation}
\noindent In the homogeneous case we find
\begin{equation}
  \label{Thomogen}
  T = T^0 = \frac{\epsilon}{k_B} \,.
\end{equation}
\noindent Therefore the homogeneous flux through the pipe
\begin{equation}
  \label{j0}
  j^0=n^0u^0=n^0\left(\frac{mg}{\gamma} -
    \frac{C~k_B T^0}{\gamma}~n^0\right)
\end{equation}
\noindent shows two distinct regimes: a low density regime 
due to a high particle velocity where only very few collisions occur
and a high density regime due to low particle velocity caused by
dissipative impacts of particles.

When we assume local equilibrium, i.e.~$n^0
\rightarrow n(x,t)$, $u^0 \rightarrow u(x,t)$, and $T^0 \rightarrow
T(x,t)$, we rewrite eq.~(\ref{FPE}):
\begin{equation}
  \label{Balance}
  \frac{\partial P}{\partial t} + \frac{\partial}{\partial x}[vP] +
  \frac{\partial}{\partial v}\left[ \left(\frac{F(n,T)}{m} - \frac{\gamma}{m}v\right)P \right] =
  \frac{\epsilon\gamma}{m^2}\frac{\partial^2 P}{\partial v^2} \,.
\end{equation}
\noindent Similar as in the Vlassov formulation~\cite{Vlassov} the force
\begin{equation}
  \label{Vlassov}
  F(n,T) = m g - C~k_B T(x,t) n(x,t)
\end{equation}
\noindent is determined to be self-consistent in its dependence on density and
granular temperature of the material. With the effective force
$F(n,T)$ acting on the particles at a given location $x$ and a
given time $t$ we find the corresponding Langevin equation for the
motion of the particles which are subjected to gravity and impacts of
other grains
\begin{eqnarray}
    \label{LangImpacts}
    \dot{x}_i &=& v_i\\ 
    m\dot{v}_i &=& -\gamma v_i + F\left(n\left(x_i,t\right), T\left(x_i,t\right)\right)+\sqrt{2\epsilon\gamma}~\xi_i(t) \nonumber\,.
\end{eqnarray}
\noindent Eqs.~(\ref{Vlassov},\ref{LangImpacts}) determine an effective 
simulation algorithm (for details see~\cite{lsg}).

 From eq.~(\ref{Balance}) we derive the hydrodynamic equations
\begin{mathletters}
\begin{eqnarray}
  \label{Hydro}
&&  \frac{\partial n}{\partial t} + \frac{\partial}{\partial
    x}[nu] = 0\\ 
&&  \frac{\partial u}{\partial t} + u\frac{\partial u}{\partial x} 
    = \frac{F(n,T)}{m} - \frac{\gamma}{m}u - \frac{k_B}{m n} \frac{\partial}{\partial x}[nT] \label{momentum} \\ 
&&  \frac{\partial T}{\partial t} + u\frac{\partial T}{\partial x} 
    = - \frac{2\gamma}{m}T + \frac{2\epsilon\gamma}{mk_B} - 
    2T\frac{\partial u}{\partial x} \label{heatbalance} \,.
\end{eqnarray}
\label{AllHydro}
\end{mathletters}
\noindent The first two terms on the right hand side of the heat balance
equation~(\ref{heatbalance}) describe the heat exchange between the
granular material and the wall, whereas the last term leads to an
effective volume viscosity.

In the approximation of quick temperature and velocity relaxation one
finds for the high damping limit ($\gamma \rightarrow \infty$) the
Burgers equation
\begin{equation}
  \label{Burgers}
  \frac{\partial n}{\partial t} + \frac{1}{\gamma}\frac{\partial}{\partial x}F
  \left( n,T^0\right) = \frac{\epsilon}{\gamma}
  \frac{\partial^2 n}{\partial x^2} \,.
\end{equation}
\noindent In this limit there are no self sustained inhomogeneous solutions of
eq.~(\ref{Burgers}) (see~\cite{Burgers}). For finite damping, $\gamma
< \infty$, however, as shown below the homogeneous solution $n = n^0$,
$u = u^0$, $T=T^0$ of the eqs.~(\ref{AllHydro}) becomes unstable when
the average density approaches a critical value $n^{cr}$. In
the context of clustering instabilities in dissipative gases
Goldhirsch and Zanetti~\cite{Goldhirsch} argued similarly:
when the pressure in a dense region decreases due to dissipation, the
resulting pressure gradient leads to further increase of the density
which finally results in a granular cluster. 

\begin{figure}[htb]
\begin{minipage}{8cm}
  \begin{center}
  \leavevmode
  \centerline{\psfig{figure=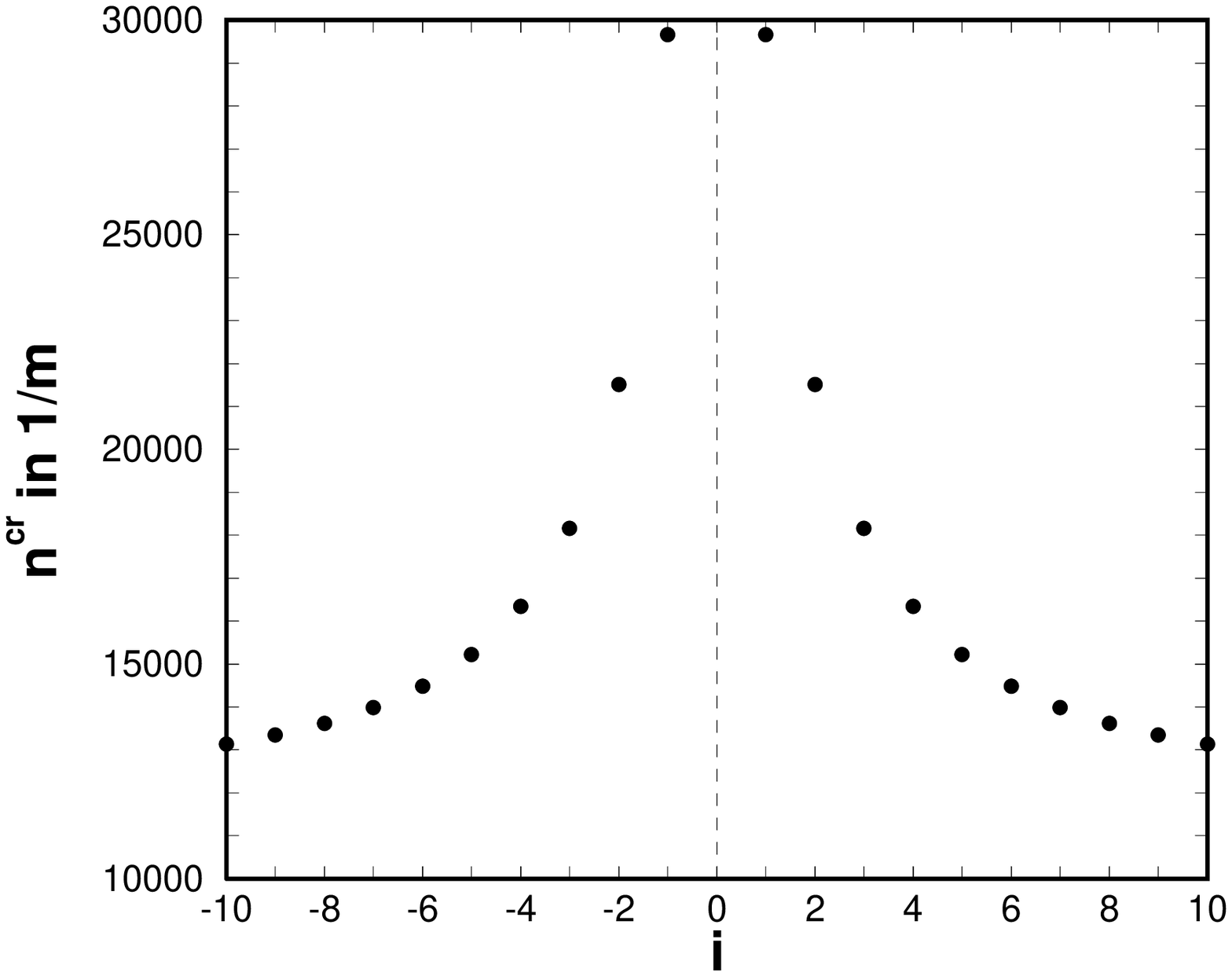,width=8cm}}
  \caption{The critical density $n^{cr}$ over the mode number $i$. 
  The parameters are $m=7.4~10^{-7}~kg$, $\gamma=7~10^{-6}~kg/sec$,
  $\epsilon=2.0~10^{-8}~Nm$, $C=6.4~10^{-3}$, $g=9.81~m/sec^2$.
  For different parameters the curve changes,
  however, its qualitative shape remains conserved.}
  \label{fig1}
\end{center}
\end{minipage}
\end{figure}

In the following we want to discuss the stability analysis of the
hydrodynamic equations (\ref{AllHydro})~\cite{Riethmueller}.
Obviously, there is a homogeneous
solution~(\ref{uhomogen},\ref{Thomogen}) for a given homogeneous
density $n^0$.  We disturb the homogeneous state in
eqs.~(\ref{AllHydro}) with wave-like perturbations ($\delta n \sim
\delta u \sim \delta T \sim \exp(-\alpha t+\mbox{i}kx)$), drop the
quadratic terms and get a eigenvalue--problem for $\alpha(k)$. If the
real part of $\alpha(k)$ is negative, $\mbox{Re}(\alpha(k))<0$,
fluctuations can grow and the initially homogeneous state becomes
unstable. The transition occurs at the critical particle density
$n^{cr}$ where $\mbox{Re}(\alpha(k))=0$.  Because of the assumed
periodic boundary conditions the wavenumber $k$ is discrete:
$k=\frac{2\pi}{L}i \quad (i=\pm 1,\pm 2,\cdots)$.  Fig.~\ref{fig1}
shows the critical density $n^{cr}$ over the dimensionless mode number
$i$.  Obviously in particular short length perturbations are able to
destabilize the homogeneous flow in the granular system. This stands
in strong contrast to results found for hydrodynamic formulations for
traffic~\cite{Prigogine,KernerKonhaeuser,KurtzeHong} and granular
flows~\cite{Savage}, where the long range fluctuations are the
critical ones.  Our results are not surprising if one imagines that
a local large gradient of the velocities will lead to a high collision
rate at this place. For sufficient high density this process leads
to clusters with high local density and small average velocity.

For large wave numbers we get a low limiting critical density given by
\begin{equation}
  \label{LimesNcr}
  \lim_{|k|\rightarrow\infty}~n^{cr}(k) = \frac{7}{3\sqrt{3}}
  \frac{\gamma}{\sqrt{\epsilon m}C} = \frac{7}{3\sqrt{3}}\frac{1}{L_B C} \, ,
\end{equation}
\noindent where $L_B=\sqrt{\epsilon~m}/\gamma$ is the braking distance, i.e.
the length a particle is damped out after an impact. It determines the
length scale which characterizes our granular system and its critical
behavior. In contrast, the critical behavior of the traffic flow
models proposed in~\cite{Prigogine,KernerKonhaeuser,KurtzeHong}
depends on the length $L$ of the entire (periodic) system too since
the critical fluctuations are long range ones.

To check the analytic results the time--discretized Langevin equations
\begin{mathletters}
  \begin{eqnarray}
    \label{LangSim}
    x_i(t + \Delta t) &=& {x}_i(t) + v_i(t)\Delta t\\ 
    v_i(t + \Delta t) &=& v_i(t) + \left( \frac{F(n,T)}{m} - 
    \frac{\gamma v_i(t)}{m}\right) \Delta t + \nonumber \\
    & & \frac{\sqrt{2\epsilon\gamma~\Delta t}}{m}\mbox{GRND}
  \end{eqnarray}
\end{mathletters}
\noindent have been solved numerically. GRND is a Gaussian random number with
standard deviation equal unity. Using the parameters
$m=7.4~10^{-7}~kg$, $\gamma=7~10^{-6}~kg/sec$,
$\epsilon=2.0~10^{-8}~Nm$, $C=6.4~10^{-3}$, $g=9.81~m/sec^2$, and $\Delta
t=0.01~sec$ we calculated the time dependent density, velocity and
granular temperature from the trajectories of the Brownian particles.
The given parameters have been determined
experimentally~\cite{Riethmueller}. We assume periodic boundary
conditions and homogeneous initial conditions. Inserting the given
parameters into eq.~(\ref{LimesNcr}) we find $n^{cr}=12000/m$.

\begin{figure}[htb]
\begin{minipage}{8cm}
  \centerline{\psfig{figure=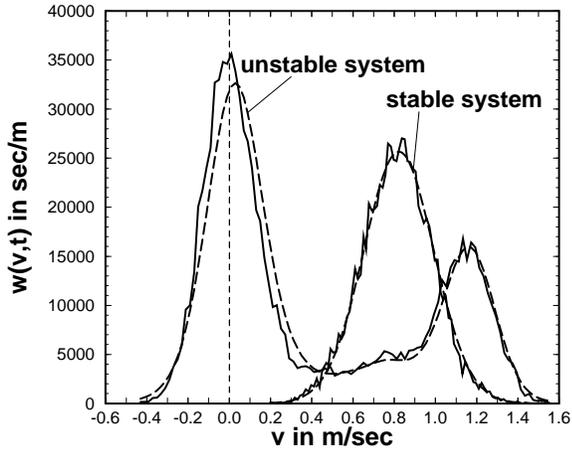,width=8cm}}
  \caption{Velocity distribution found by simulations of the stable 
  homogeneous and the unstable system (solid lines) and the
  corresponding analytical results (dashed lines).}
  \label{fig2}
\end{minipage}
\end{figure}

Fig.~\ref{fig2} shows the velocity distributions 
\begin{equation}
  \label{VeloDistr}
  w(v,t) = \int_{0}^{L} P(x,v,t)dx
\end{equation}
\noindent of a stable (undercritical) ($n^0=11000/m<n^{cr}$) and an unstable
system ($n^0=14000/m>n^{cr}$). In the undercritical case we find a
stable homogeneous flow with Gaussian velocity distribution, while in
the latter case inhomogeneities due to random fluctuations increase
with time and the velocity distribution $w(v,t)$ is no longer
Gaussian. In our opinion there are at least two distinct velocity
distributions in the system: at regions of low density we find high
average grain velocity and at high density regions the grains move
with low average velocity. The overlayed curves in the distribution
plots (fig.~\ref{fig2}) show support of the assumption of a bimodal
velocity distribution.

\begin{figure}[htb]
\begin{minipage}{8cm}
  \centerline{\psfig{figure=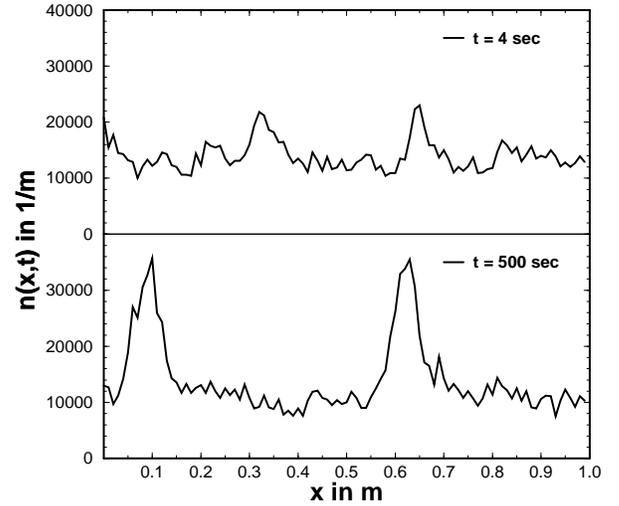,width=8cm}}
  \caption{The particle density of the unstable system at times 
  $t= 4sec$ and $t= 500sec$. Since the initial (homogeneous) density
  is overcritical the inhomogeneities increase with time.}
  \label{fig3}
\end{minipage}
\end{figure}

\begin{figure}[htb]
\begin{minipage}{8cm}
  \centerline{\psfig{figure=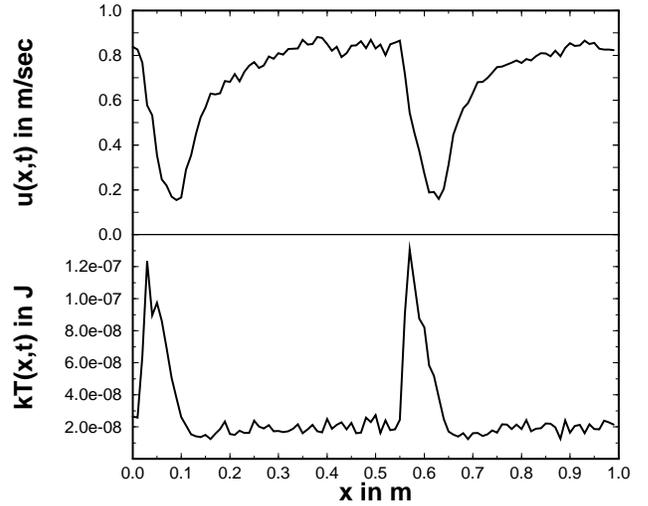,width=8cm}}
  \caption{The mean
  velocities (top) and the mean square displacement of the velocity
  (bottom) of the unstable system at time $t=500sec$.} 
  \label{fig4}
\end{minipage} 
\end{figure}

Fig.~\ref{fig3} shows snapshots of the particle density of the
unstable system where gravity acts in positive $x$-direction.  We
eventually observe the formation of two moving clusters originating
from random inhomogeneities in the early state of the
system--evolution.  In dependence from the initial conditions we found
also configurations with one or three moving clusters. In
correspondance with the experiment and MD simulations~\cite{Poeschel}
we observed coexisting clusters moving either in positive or negative
direction. Note that the slope at the left hand side of the density
wave in fig.~\ref{fig3} is very steep. Here the collision rate is very
high due to the large velocity gradient between the particles which
are involved in the clusters and the free falling ones
(fig.~\ref{fig4}). The particle velocity at the right hand side of the
clusters is much lower. There the grains' velocity slowly increases
under the influence of gravity and hence the high density area, i.e.
the cluster, dissolves (see also~\cite{DuranLuding}).  These processes
lead to the obvious hump-like shape of the clusters in
fig.~\ref{fig3}. Contrary to the hump-like solutions of the
Burgers-equation~\cite{Burgers} the widths of the clusters remains
invariant when they move through the pipe. Fig.~\ref{fig4} shows the
described sharp decrease and slowly increase of the particle
velocities. As one can derive from the hydrodynamic
equations~(\ref{heatbalance}) the high negative velocity gradient at
the front (left hand side) of a cluster leads to an increase of the
granular temperature, whereas the small positive velocity gradient
inside and at the backside of the cluster results in a smaller
granular temperature compared with outside the clog. Fig.~\ref{fig4}
(lower part) illustrates this plausible behavior.

We investigated the granular flow trough a vertical narrow pipe using
a simple model consisting of Brownian particles with collision
interaction. We could show that there is a critical value for the
particle density which decides whether a initially homogeneous flow
remains stable. The model is valid in the limit of pairwise particle
interaction. This precondition is assumed to be fulfilled for the case
of moderate particle density and low pipe width.  Our model also
defines an efficient simulation method for granular flows.  Applying
this algorithm to low- and high-density pipe flow we found that in
both cases the numerical results agree with the theoretical
prediction. The numerical results for the spacial particle density,
the average velocity, and the granular temperature agree with the
hydrodynamic description.

We thank S.~Esipov, S.~Savage, J.~Schuchhardt, and H.~J.~Herrman for
useful discussion.

\newpage

\end{multicols} 
\end{document}